\documentclass{eas}
\usepackage{graphicx}
\newcommand{\bm}[1]{\mbox{\boldmath $#1$}}

\def\be{\begin{equation}}
\def\ee{\end{equation}}
\def\bea{\begin{eqnarray}}
\def\eea{\end{eqnarray}}
\def\bean{\begin{eqnarray*}}
\def\eean{\end{eqnarray*}}

\newtheorem{theorem}{Theorem}

\begin{document}

\title{A New Type of Singularity Theorem} 
\author{Jos\'e M M Senovilla}\address{F\'{\i}sica Te\'orica, Universidad del Pa\'{\i}s Vasco, 
Apartado 644, 48080 Bilbao, Spain}
%
%
\begin{abstract}
A new type of singularity theorem, based on spatial averages of physical quantities, is presented and discussed. Alternatively, the results inform us of when a spacetime can be singularity-free. This theorem provides a decisive observational difference between singular and non-singular, globally hyperbolic, open cosmological models.
\end{abstract}
\maketitle
\section{Introduction}
In 1990 I presented the following line-element (Senovilla \cite{S})
\begin{eqnarray}
ds^2=\cosh^4(at)\cosh^2(3a\rho)(-dt^2+d\rho^2)+\hspace{2cm}\nonumber\\
\frac{1}{9a^2}\cosh^4(at)\cosh^{-2/3}(3a\rho)\sinh^2(3a\rho)d\varphi^2+
\cosh^{-2}(at)\cosh^{-2/3}(3a\rho)dz^2, \label{sol}
\end{eqnarray}
given in cylindrical coordinates $\{t,\rho,\varphi,z\}$, $a>0$ is a constant. It is a {\em cylindrically symmetric} perfect-fluid solution of the Einstein field equations. The energy density $\varrho$ and the unit velocity vector field $\vec u$ of the fluid are given by
$$
\varrho = 15a^2\cosh^{-4}(at)\cosh^{-4}(3a\rho) \, ,\hspace{4mm}
\vec u=\cosh^{-2}(at)\cosh^{-1}(3a\rho)\, \partial_{t}
$$
while the acceleration 
one-form is non-zero $\bm{a}=3a\tanh (3a\rho)d\rho$.
The fluid satisfies the  barotropic equation of state ($p$ is the isotropic pressure)
$$
p=\frac{1}{3}\varrho
$$
which is {\em realistic} for radiation-dominated matter, that is, for the initial stages of the present expanding era of the Universe.

The space-time (\ref{sol}) satisfies the strongest causality condition,
global hyperbolicity, as any $t=$const.\ slice is a Cauchy hypersurface. Observe that $\varrho$ and $p$ are regular and positive everywhere. Actually, 
the space-time (\ref{sol}) is completely free of 
singularities and geodesically complete. For thorough discussions see (Chinea \etal \cite{CFS}; Ruiz and Senovilla \cite{RS}; Senovilla \cite{S2} sect.~7.6). Given that $p>0$ the solution appears to be ``cosmological'', as it cannot be matched to any vacuum exterior and matter is present everywhere. The fluid expansion reads
$$
\theta=\nabla_{\mu}u^{\mu} = 3a\frac{\sinh(at)}{\cosh^3(at)\cosh(3a\rho)}
$$
so that the entire universe is contracting {\em 
everywhere} if $t<0$ 
and expanding if $t>0$, having a rebound at $t=0$. Thus, this simple model showed that there exist well-behaved 
classical models expanding everywhere, satisfying all energy and 
causality conditions, and singularity-free. However, (\ref{sol}) is an {\em unrealistic} model to describe the actual Universe because the cylindrical symmetry implies a type of anisotropy incompatible with observations (Senovilla \cite{S1}).

The following questions spring to mind: how many well-founded non-singular solutions are there? did the universe have the choice of a non-singular start of the expansion epoch? As a matter of fact Garfinkle \& Harris (\cite{GH}) answered the related question of how the Ricci tensor behaves in {\em stationary} globally hyperbolic geodesically complete spacetimes. However, the interesting case is {\em non-stationary}, of course. This is the case treated in (Senovilla \cite{Snew}) upon which this contribution is based. The following is a brief summary of the results.
\section{Averaging}
In 1998, Raychaudhuri (\cite{R2}) proved that the solution (\ref{sol}), as well as the many other singularity-free solutions found since 1990 ---see (Senovilla \cite{S2,Snew}) for lists and a review---, has {\em vanishing space-time averages of the energy density and pressure}. However, it was soon realized (Saa \cite{Saa}; Senovilla \cite{S3}) that this same property is shared by all Robertson-Walker models. Hence, these averages do not distinguish between singular and non-singular cosmological models. 

Nevertheless, Raychaudhuri was pointing into a very interesting direction: averaging. The use of pure spatial averages (at a given instant of time) occurred to me immediately, and a conjecture was put forward in (Senovilla \cite{S3}). After a hazardous life, see e.g. the chronologically key references (Raychaudhuri \cite{R3,R4,R5,R6}; Fern\'andez-Jambrina \& Gonz\'alez-Romero \cite{LeoMan1,LeoMan2}; Fern\'andez-Jambrina \cite{Leo2}), the conjecture was made precise and recently proven in (Senovilla \cite{Snew}; Jerjen \cite{J}). To understand its rigorous formulation we need to learn a little about spatial averages.

\subsection{Spatial averages} 
Let $\Sigma$ be any spacelike hypersurface and 
$\bm{\eta}_{\Sigma}$ its canonical volume element 3-form. The average $\left<f\right>_S$ of any scalar $f$ on a 
finite portion $S$ of $\Sigma$ is defined by
$$
 \left<f\right>_{S}\equiv 
 \frac{\displaystyle{\int_{S}f\bm{\eta}_{\Sigma}}}
{\displaystyle{\int_{S}\bm{\eta}_{\Sigma}}}=
[{\rm Vol}(S)]^{-1}\int_{S}f\bm{\eta}_{\Sigma}
$$
where ${\rm Vol}(S)$ is the volume of $S\subseteq \Sigma$. Let $\{S_{i}\}$ be a continuous sequence of nested such portions converging to $\Sigma$ ---for the notion of convergence, see (Jerjen \cite{J}).
The spatial average on the whole $\Sigma$ is defined then as: 
\begin{equation}
 \left<f\right>_{\Sigma}\equiv 
 \lim_{i\rightarrow \infty}
[{\rm Vol}(S_{i})]^{-1}\int_{S_{i}}f\bm{\eta}_{\Sigma}
\label{average}
\end{equation}
This definition is independent of the chosen sequence $\{S_{i}\}$ (Jerjen \cite{J}).

The averages (\ref{average}) satisfy many useful properties (Senovilla \cite{Snew}), but the crucial one is that if $f\geq 0$ on $\Sigma$, then
$\left<f\right>_{S}\geq 0$ and the equality 
requires necessarily that $f\rightarrow 0$ along {\em almost every} direction ``approaching the boundary'' (i.e., 
when going to infinity). Conversely, if $f> 0$, $f$ is bounded on $\Sigma$, $f$ is asymptotically non-oscillatory (Jerjen \cite{J}), 
and $f$ is bounded from below by a positive constant at most along a set of directions of zero measure, 
then $\left<f\right>_{S}= 0$. The non-oscillatory behaviour introduced in (Jerjen \cite{J}) to correct the proofs in (Senovilla \cite{Snew}) is perhaps too strong a requirement, but I have not been able to find a milder restriction hitherto.

\section{The Main Theorem}
The conjecture put forward in (Senovilla \cite{S3}) was promoted to an actual singularity theorem in (Senovilla \cite{Snew}). There were some technical details, however, which made the proof incomplete. These have been corrected in (Jerjen \cite{J}) while keeping the spirit and the body of the original demonstration. 

The main theorem reads:
\begin{theorem}
Assume that 
\begin{enumerate}
\item the spacetime contains a non-compact Cauchy hypersurface $\Sigma$ such that its 
second fundamental form has a trace positive everywhere and asymptotically non-oscillatory
\item $R_{\rho\nu}v^{\rho}v^{\nu}\geq 0$, where $\vec v$ is the geodesic vector field  orthogonal to $\Sigma$ on $\Sigma$.
\item the scalar curvature $\overline{R}$ of $\Sigma$ is non-positive on average on $\Sigma$: $\left<\overline{R}\right>_{\Sigma}\leq 0$;
\item the cosmological constant is non-negative $\Lambda \geq 0$; and
\item the dominant energy condition holds.
\end{enumerate}
If any single one of the following spatial averages
\begin{eqnarray*}
\Lambda , \,\, \left< \theta \right>_{\Sigma}, \,\, \left< \vartheta \right>_{\Sigma} ,
\,\, \left< \theta^2 \right>_{\Sigma}, \,\, \left< \vartheta^2 \right>_{\Sigma} ,\,\,
\left<K_{\mu\nu}K^{\mu\nu}\right>_{\Sigma},Ê\,\, \left<\overline{R}\right>_{\Sigma},\\
\left<v^{\mu}\nabla_{\mu}\vartheta\right>_{\Sigma},\, \,
\left<u^{\mu}\nabla_{\mu}\theta-\nabla_{\mu}a^{\mu}\right>_{\Sigma},\hspace{15mm}\\
\left<T_{\mu\nu}e^{\mu}_{\alpha}e^{\nu}_{\beta}\right>_{\Sigma}, \hspace{3mm}
\left<R_{\mu\nu}e^{\mu}_{\alpha}e^{\nu}_{\beta}\right>_{\Sigma} \hspace{3mm}
\forall \alpha,\beta =0,1,2,3
\end{eqnarray*}
does \underline{not} vanish, then {\em all} past-directed timelike geodesics are
incomplete.
\label{th4}
\end{theorem}
Here $R_{\mu\nu}$ is the Ricci tensor, $\vartheta=\nabla_{\mu}v^{\mu}$ is the expansion of $\vec v$, while $\vec u$ is {\em any} unit timelike vector field such that $\vec u|_{\Sigma}=\vec v|_{\Sigma}$, $\theta$ is its expansion and $\vec a=\nabla_{\vec u}\vec u$ its acceleration vector field. $\{\vec{e}_{\alpha}\}$ is any orthonormal basis. Finally, $K_{\mu\nu}$ is the second fundamental form of $\Sigma$, which coincides on $\Sigma$ with the shear tensor of $\vec v$ (and equals the shear tensor of $\vec u$ minus the symmetrized tensor product of $\vec a$ and $\vec u$.) Of course, $\vartheta =K^{\mu}{}_{\mu}$ so that $\vec v$ is expanding everywhere on $\Sigma$, and the expansion is asymptotically non-oscillatory on $\Sigma$.

Let me briefly comment on the reasonability of the assumptions and the strength and meaning of the derived result. First of all one requires the  
spacetime to be globally hyperbolic, so that it is causally 
well-behaved. Furthermore, the Universe is assumed to be {\em open}
($\Sigma$ is non-compact) and everywhere expanding 
at a given instant of time (say ``now''), described by the hypersurface $\Sigma$.
The timelike convergence condition must also hold along the 
geodesic congruence orthogonal to $\Sigma$, and the dominant energy condition is added. All this is standard and well motivated and founded.

On the other hand, the {\em space} of the Universe (represented by $\Sigma$ at 
the expanding instant) is assumed to be non-positively curved {\em on average}, while $\Lambda$ is taken as non-positive. Observe that the traditional case with 
$\Lambda =0$ is included. Both of these requirements are in agreement with present observational data, e.g. the recent data 
from WMAP. The assumption on $\overline{R}$ still allows, actually, for an everywhere positively curved $\Sigma$. As a matter of fact, the solution (\ref{sol}) has $\overline{R}>0$ everywhere (Senovilla \cite{Snew}).

Given the above, the conclusion of the theorem can be stated in two alternative, but equivalent, ways: (i) in order to have past geodesic completeness, {\em all} the displayed spatial averages must vanish, including in particular those of the energy density, the pressure, and all physical components of the energy-momentum tensor; alternatively, (ii) if any of these averages is non-zero, the Universe must be {\em totally} past geodesically incomplete. The universality of the past incompleteness is a direct consequence of the technical assumption that $\vartheta$ is asymptotically non-oscillatory, which implies a kind of uniformity for the limit when going to infinity along $\Sigma$ (Jerjen \cite{J}).

The proof of the theorem has three fundamental ingredients: the Raychaudhuri (\cite{R}, \cite{R1}) equation  ---see also (Hawking \& Ellis \cite{HE}; Wald \cite{Wald})---, the simplest standard singularity theorem (Hawking \& Ellis \cite{HE}; Penrose \cite{P1}; Senovilla \cite{S2}), and the so-called Hamiltonian constraint, e.g. section 10.2 in (Wald \cite{Wald}). For details, check (Senovilla \cite{Snew}; Jerjen \cite{J}).

\section{Conclusions}
The theorem has the following fundamental implication: under the
stated assumptions, a non-vanishing average of {\em any} component
of the energy-momentum tensor, or of any kinematical variable, leads
to the existence of a kind of big-bang singularity in the past. Observe
that there is no need to assume any specific type of
matter content (such as a perfect fluid, scalar field, etc.), as only the
physically compelling dominant energy condition
is required. This is remarkable.

The theorem concerns ``open" models, as the Cauchy hypersurfaces are non-compact. This is, however, not a relevant limitation
because there are stronger
results proving the geodesic incompleteness of closed models (Hawking \& Ellis \cite{HE}; Hawking \& Penrose \cite{HP}; Penrose \cite{P,P1}; Senovilla \cite{S2}). As a matter of fact, closed expanding
non-singular models require the violation of the timelike convergence condition $R_{\mu\nu}v^{\mu}v^{\nu}\geq 0$ (Senovilla \cite{S2}) ---without this condition, but satisfying the dominant energy condition, there are some non-singular acceptable examples (Mars \& Senovilla \cite{MS}; Senovilla\cite{S2}). 

I would like to stress that the conclusion in theorem \ref{th4} is very strong:
it tells us that the incompleteness is {\em to the past}, and for the
{\em entire} class of timelike geodesics.  
Besides, I believe that one can in fact do better and get a stronger theorem such that the non-oscillatory condition on the expansion can be substantially relaxed, or even removed. 
 
Our theorem states, in fact, that a clear, decisive,
difference between singular and regular, globally hyperbolic,
everywhere expanding (with non-oscillatory behaviour at infinity) cosmological models is that the former can have non-vanishing
spatial averages of some of the matter variables, a property which is forbidden for the latter. This may seem to imply that {\em regular} globally hyperbolic expanding models {\em cannot} be of cosmological type and realistic, as long as the Universe contains a sufficiently homogenous distribution of matter ---e.g. not decaying away from us. 
 
On the whole, this is a very satisfactory conclusion. 

\section*{Acknowledgments}
Comments from Harald Jerjen are acknowledged.
Support under grants FIS2004-01626 (MEC) and GIU06/37 of the University of the Basque Country (UPV/EHU) is gratefully acknowledged.


\end{document}